# Photostrictive Effect and Structure Phase Transition via Nonlinear Photocurrent

Ruixiang Fei [1, 2, *] and Li Yang [2, 3, †]

[1] *School of Physics, Key Lab of Advanced Optoelectronic Quantum Architecture and Measurement (Ministry of Education), Beijing Institute of Technology, 100081, Beijing, China*

[2] *Department of Physics, Washington University in St Louis, St Louis, Missouri 63130, United States*

[3] *Institute of Materials Science and Engineering, Washington University in St. Louis, St. Louis, Missouri 63130, United States*

Abstract:

The phenomena of crystal size changes and structural phase transitions induced by light irradiation have garnered significant interest due to their potential for tuning and controlling a wide range of material properties through highly cooperative interactions. However, a microscopic theory that can comprehensively explain these phenomena in correlation with photon frequency and polarization has remained highly desirable. In this work, we propose that nonlinear photocurrent may correspond to driving these effects, which arise from a competition between light-injected energy and structural variations. By conducting first-principles simulations and comparing them with two established experiments, we show that shift current, a second-order photocurrent, can induce photostriction and nonreciprocal structure phase transitions. The quantitative comparisons across key parameters such as light frequency, irradiation time, polarization, and intensity provide further support for the nonlinear photocurrent mechanism. Beyond shift current, this microscopic understanding proposes to utilize more types of nonlinear photocurrent to enhance light-structure interactions and control material properties.

***Introduction***: To date numerous exotic light-induced phenomena in matter have been discovered [1], sparking significant interest because of their potential for manipulating and controlling exotic quantum states. Notable examples include light-induced superconductivity in stripe-ordered Cuprates [2], photon-driven phase transition in $VO_2$ [3] and ferroelectrics [4–6], and light-induced anomalous Hall effect [7,8]. Among these phenomena, the interactions between light and structures have garnered particular attention, as structural variation can impact a wide range of fundamental properties. For instance, extensive studies have demonstrated substantial photostriction in noncentrosymmetric semiconductors [9–12], and phase transitions associated with topological states have been observed in transition metal dichalcogenides (TMDs) under above-band gap laser irradiation [13–15].

These light-induced phenomena can often be understood using perturbation methods. One notable approach is the Floquet-Bloch band theory, which provides insight into the light-induced topological band transition [16–19]. However, it is challenging to understand photostriction and structural phase transitions using the Floquet-band theory, which does not correspond to the lattice information. The ultrafast phase transition induced by Thz laser can be replicated using real-time time-dependent density functional theory [20–22], where the outcomes depend on a chosen excited exchange-correlate potential [23]. However, explaining the persistent photostriction and phase transitions caused by the above-band-gap laser remains challenging. Recently, the constrained density functional theory (DFT) [24,25] has been employed to investigate photostriction and phase transitions in ferroelectrics. This approach utilizes density matrix constraints to control the orbital occupation numbers and provides insights into the behavior of excited states. Notably, the intensity of illumination can be converted to the number of photoexcited electron-hole pairs [26], although the light-frequency dependence is not included. Nonetheless, the quest for a microscopic theory that can fully account for the observed photostriction or phase transition, while considering the correlation with photon polarization and energy [9,10,27], remains desirable.

We note several key characteristics that are crucial for observing those significant light-structure interactions. Firstly, many of these phenomena have been observed under strong light-intensity conditions, indicating the emergence of the nonlinear optical response such as shift current, inject current, and second-harmonic generation in ferroelectrics [28–31], CdS [32], and TMDs [33]. Secondly, photostriction tends to be more pronounced in noncentrosymmetric materials. For example, the measured amplitude of photostriction of CdS [9] and GaAs [34] is about three orders of magnitude larger than that of nonpolar Germanium [35], indicating the significant role of second-order light-matter interactions, which require the broken inversion symmetry.

In this work, we propose that the photostricive effect and photo-driven structure phase transitions are stimulated by photocurrent originating from nonlinear optical effects. To validate our proposal, we first conducted a case study using the polar CdS semiconductor. Our results demonstrate a remarkable consistency between the calculated photostriction and experimental data. Additionally, we reveal that this mechanism can drive an observed irreversible structural phase transition in layered $MoS_2$, transforming it from the 2H phase to the 1T' phase, which is known to exhibit the topological-insulator behavior. These agreements with experiments support that nonlinear photocurrent may be a universal explanation for photostriction and light-driven structural phase transitions in polar materials. Based on this understanding, we can select appropriate light polarization and symmetries of matter to enhance nonlinear photocurrent and subsequent light-structural interactions.

***Energy barrier eliminated by photocurrent or photovoltage***: In a homogenous insulator under above-band-gap irradiation, excited carriers can be categorized into three phases: photoexcitation, thermalization, and recombination. However, the current $\frac{d\mathbf{P}}{dt}$, where $\mathbf{P}$ represents the electric polarization, can be nonzero due to high-order light-matter interactions. In a phenomenological scenario shown in Fig. 1(a), this can cause the excited carriers, i.e., electrons and holes, to move to the opposite sides of the material before they recombine. Consequently, the accumulated charge forms a capacitor under the open-circuit situation, as depicted in Fig. 1(b). Thus, the leading-order energy gain from the light in homogeneous material can be expressed as

$$\Delta E = \frac{1}{2}QV \tag{1}$$

where $Q = \int_0^t I(\tau)d\tau$ and $V = \frac{1}{C}\int_0^t I(\tau)d\tau$ refer to the photoinduced accumulated charge and voltage, respectively. $I$ is the nonlinear photocurrent, $C$ is the capacitance, and $t$ represents the irradiation time. Because the direction and amplitude of nonlinear photocurrent are sensitive to the atomic structure, the capacitor model may acquire different amounts of energy for different structures, causing the potential to shift from the black line to the red line, as shown in Fig. 1(c). This increase in potential energy may lead to the metastable structure being a more stable configuration, ultimately resulting in a structural phase transition. Such an occurrence of the light-induced structural phase transition depends on several factors such as potential barrier and characteristics of the metastable structure. Nonetheless, the change in potential energy caused by light can generally alter the deformation of the material, which is referred to as the photostrictive effect.

There are numerous types of second-order photocurrents, such as the shift current [28,36,37], injection current [37,38], magnetic injection current [39,40], ballistic current [41], and gyration current [42,43]. Given that the photocurrent observed in our compared experiments is mainly created by the ambient light, which can be regarded as a superposition of randomly linearly polarized light, we concentrate on the shift current because it is generated by linearly polarized light in nonmagnetic materials.

Shift current density $J_c = \sigma_{ab}^c E_a(\omega) E_b(-\omega)$ mainly describes the polarization difference between the conduction band and valence band [36,37,44,45], where $E(\omega)$ is the electrical field of incident light with a frequency $\omega$, indices $a$, $b$, and $c$ are the real-space directions. $\sigma_{ab}^c$ is the second-order current susceptibility in the form of

$$\sigma_{ab}^c = \frac{i\pi e^3}{\hbar^2 \omega^2} \sum_{mn} \int d^3k (v_{mn}^a(k) v_{nm;c}^b(k) + v_{mn}^b(k) v_{nm;c}^a(k)) \delta(\omega - \omega_{mn}) \quad (2)$$

where $v_{mn}^b$ is the velocity matrix element, $v_{nm;c}^b \equiv \frac{\partial v_{nm}^b}{\partial k^c} - i[A_{nn}^c - A_{mm}^c] v_{nm}^b$ is the gauge-independent "generalized derivatives" of the velocity matrix element, and $A_{nn}^c$ is the Berry connection of the band $n$. In our first-principles calculations, we adopt two methodologies for the shift current calculation, *i.e.* the sum rule $v_{nm;c}^b = -\frac{v_{nm}^b(v_{nn}^c - v_{mm}^c)}{\omega_{nm}} - \sum_l \left( \frac{v_{nl}^b v_{lm}^c}{\omega_{nl}} - \frac{v_{nl}^c v_{lm}^b}{\omega_{lm}} \right)$ with a significant number of unoccupied band $l$, and the other so-call shift vector method for validating our results [46].

*Photostricive effect in CdS*: The photostrictive effect, which involves light-induced dimension changes, is commonly observed in non-centrosymmetric materials, such as CdS, GaAs, and BiFeO$_3$ [9,10,12,27,34]. Without sacrificing generality, we compare the calculated results with the seminal photostriction observed in a nonmagnetic crystal CdS [9]. We employ first-principles simulations (see Supplemental Material [46–50]) to calculate shift current and photostriction in bulk CdS. The calculations were performed using experimental geometry at room temperature [9]. Under such a situation, CdS crystallizes in the wurtzite lattice [space group $P6_3mc$ ($C_{6v}^4$)], which is also the ground state using first-principles calculation, as shown in the inset of Figure 2 (a). Due to its space group symmetry, the shift current susceptibility satisfies $\sigma_{xx}^z = \sigma_{yy}^z$ (see Fig. S2 in Supplemental Material [46]). This implies that the z-axis (i.e. [001]) direction currents induced by the x-axis (i.e., [100]) direction polarized light and y-axis (i.e., [120]) direction polarized light are identical. Thus, the photo-induced current along the z-axis direction is independent of light polarization.

Using the principle of minimum energy, we determine the photostriction based on the sum of lattice change-induced energy variation and photon-injected energy (charging energy). If we simplify Eq.(1) and estimate $\int_0^t I(\tau)d\tau = \frac{1}{2}It$, the photo-injected energy can be written as $\Delta E = \frac{1}{8C}I^2t^2$, where the illustration time is set to $t = 1s$. The parallel-plate capacitance $C = \frac{\epsilon_0 \epsilon_r l \cdot w}{d}$ is linked to the sample dimensions from Ref. [9]: length $l = 10\ mm$, width $w = 3\ mm$, and depth $d = 0.015\ mm$. The absolute permittivity of CdS at room temperature along the [001] direction is set to $\epsilon_r = 10.16$ from the handbook [51]. Because the sample ($d = 15\mu m$) is thinner than the light penetration depth (a few hundred $\mu m$), the total current for polarization perpendicular to the z-axis can be computed by $I = \sigma_{ab}^c \cdot l \cdot w \cdot E_s^2$, where $E_s$ is the light electric field which is determined from the irradiation intensity $I_s = 7.5\ mW/cm^2$, which is based on H. C. Gatos et.al. seminal experiments [9,34]. Remarkably, we find that the light-injected energy per unit cell $\Delta \mathcal{E} = \frac{(\sigma_{ab}^c E_s^2 t)^2 a_1 a_2 a_3}{8\epsilon_0 \epsilon_r}$ is irrelevant to the sample dimensions, where $a_1, a_2$, and $a_3$ are the unit cell lattice constants of wurtzite CdS (see Supplemental Material [46])

Figure 2(a) shows the total energy vs the lattice constant (in the z-axis direction) of intrinsic wurtzite CdS. Different curves respond to different incident photon energies, and the minimum of each curve decides the stable lattice constant under that specific-frequency irradiation. To align with the experimental optical band gap (2.42 eV), our first-principles DFT bandgap and photon energy are shifted by 1.31 eV. As shown in Fig. 2 (a), wurtzite CdS exhibits a photostriction for all considered light frequencies. For example, under 2.52-eV irradiation, the lattice contracts about 1.2%, as demonstrated by the blue dash-dotted line. The reason for the contraction is that the shift current of a stretched structure is greater than those of the intrinsic or compressed structures (see Supplemental Material [46]). Thus, the total energy of the stretched structure increases more under irradiation, which is not preferred by the principle of minimum energy.

We also compare the calculated photostriction spectrum with the experimental measurement [9], as plotted in Fig. 2(b). The overall agreement is very good. Moreover, the spectrum profile is insensitive to light intensity $I_s$ and/or irradiation time $t$. The profile in Fig. 2(b) remains close to the measured results for different light intensity or irradiation time (See Fig. S5 in Supplemental Material [46]). These robust results validate our proposal, despite the measured spectrum in Ref. [18] is in arbitrary units. To further check these results, we have also employed the shift vector method with different software packages and obtained similar outcomes (see Fig. S4 in Supplemental Material [46]). It is worth mentioning that the experiment observed a small shoulder below the band gap, as denoted by the red line close to the band edge. This can be attributed to the excitonic effect, which is not considered in our first-principles calculation.

In the above calculation, an assumption is that most light-injected energy transforms to charge the material through the parallel-plate capacitor model. We have calculated the heat energy generated by the photocurrent. Interestingly, with the resistivity $\rho = 1 \Omega \cdot cm$ of this CdS sample reported from the experiment [9], the heat induced by photocurrent is $\mathcal{E}_h = I^2 R t$. We find it is about three to four orders of magnitude smaller than the 'capacitor' energy $\Delta \mathcal{E}$. Therefore, the nonlinear photocurrent-induced injected energy by the "capacitor" or polarization effect is the leading order for the photostrictive effect.

***Structural phase transition in TMDs:*** It has recently been reported that laser radiation can cause TMDs to shift from the 2H phase to the 1T' phase [15,52], which manifests the quantum spin Hall effect [53]. Fig. 3(a) presents the first-principles nudged-elastic band (NEB) calculated energy potential of MoS$_2$ from the 2H phase to the 1T' phase, and it agrees with the results reported in Ref. [53]. Following this path, the upper layer of S atoms (blue atoms) shifts upwards, causing the structural transition from the 2H phase to the nearly degenerate 1T' or 1T phase. However, the NEB result in Fig. 3 (a) indicates that realizing this structural transition is difficult because of the ~1.5-eV energy barrier. Given that such a transition has been observed in numerous experiments on TMDs under above band-gap laser irradiation [13–15], the underlying physical mechanism remains unclear.

Figures 3(b) and 3(c) show the calculated shift current susceptibility of the 2H and 1T' phases of monolayer MoS$_2$, respectively. The shift current susceptibility of the 2H phase is about one or two orders of magnitude larger than that of the 1T' phase. Consequently, when the 2H phase is illuminated with light, its total energy increases faster than the 1T' phase, ultimately triggering the potential phase transition.

In the following, we calculate light-injected energy based on shift current by using Eq. (1): $\Delta E = \frac{1}{2}QV$. Because we find that this structure phase transition undergoes through metallic structures, which results in a significant change of the dielectric constant, we estimate the averaged light-injected energy by $\Delta E = \frac{1}{4}IVt$. The photovoltage of the laser-beam patterning can be estimated by $V = E_r l$, where $l$ is the length of the patterning channel, the effective photovoltage $E_r \sim 0.2 \text{mV}/\mu\text{m}$ is obtained from the observed photovoltage in Ref. [54]. Thus, the light-injected energy in a unit cell is $\Delta \mathcal{E} = \frac{1}{4}\sigma E_s^2 E_r \cdot t \cdot a_1 \cdot a_2 \cdot a_3$, (see Supplemental Material [46]), where $a_1, a_2$, and $a_3$ represent the lattice constants. The effective thickness of a monolayer is set to $a_3 = 5\text{Å}$, and the irradiation time is set to $t = 10\text{s}$ based on Ref. [14]. The electrical field $E_s$ of incident light is gotten from the laser intensity of Ref. [15]. Like the case of CdS, $\Delta \mathcal{E}$ is irrelevant to the patterning geometry. With these parameters, we can obtain the energy of the transition NEB path that is illustrated in Fig. 3(a). Unfortunately, we find that this NEB path is not the correct one for the photon-induced phase transition. Because the shift current of the intermediate "TS" structure is greater than that of the 2H phase, light irradiation increases the energy of the "TS" structure more significantly than that of the 2H phase. As a result, light irradiation will increase the NEB energy barrier, prohibiting the structural phase transition.

Fortunately, we have identified another possible reaction path, as depicted in Figs. 3(d) and 3(e). The reaction path comprises two stages. The first stage involves the upward movement of both up-layer and down-layer S atoms, as depicted by the IV structure, leading to the eventual formation of a rectangular lattice with inversion symmetry. In the second stage, both up-layer and down-layer S atoms move in opposite directions simultaneously, maintaining the inversion symmetry, as shown by the VII structure. The energy barrier associated with this reaction path is about 3eV with no light irradiation, and it is higher than the NEB barrier. However, under light irradiation, the intermediate structures (IV and VII) keep inversion symmetry, meaning zero shift current and zero gained energy from light. Consequently, the light irradiation relatively increases the energy of the 2H phase, ultimately eliminating the transition barrier when irradiated by $1\mu m$ diameter laser beam with $15mW$ power, as illustrated in Fig. 3(d). It is worth noting that, unlike the 1T' or 2H phase, the 1T phase keeps the inversion symmetry, and its relative energy is not influenced by irradiation, as shown in Fig. 3(d). It is worth mentioning that this reaction path in Fig. 3(e) may not be the most optimized one. However, it can be regarded as an upper limit for the photo-induced structural phase transition.

Finally, despite using a higher intensity of light, the reverse phase transition from the 1T' to the 2H phase was not observed in the experiment [13]. This implies that such a nonreciprocal structure phase transition cannot be solely attributed to the photo-induced doping or laser-induced heat mechanism. Conversely, our proposed second-order photocurrent mechanism can effectively account for this nonreciprocal phase transition, as shown in Fig. 3(d).

*Outlook*: In general, this mechanism offers an explanation as to why the photostriction is more easily detected in noncentrosymmetric materials. For example, sizable photostriction is observed in polar $BiFeO_3$ even upon exposure to a bulb [10], and photostriction in nonpolar germanium is three orders of magnitude smaller than that in polar GaAs semiconductors [34,35]. The reason is that the second-order optical response can generate photocurrent in noncentrosymmetric materials [37] but only third-order or higher-order light-matter interactions can generate much weaker photocurrent in materials with centrosymmetry [45]. Beyond the discussed cases based on shift current, the injection current, another important type of second-order photocurrent, is usually more than an order of magnitude larger than the shift current [39,40,55]. Thus, those light-induced phenomena may be further enhanced by using circularly polarized light irradiation on non-magnetic materials, in which the inject current is excited. Additionally, our proposed mechanism may also apply to the light-induced antiferromagnetic-to-ferromagnetic phase transition [56] by considering magnetic injection photocurrent [39,40,57] and gyration current [42], and it could potentially explain light-induced structural phase transitions in metals [58,59] by the interband photocurrents, rectification, and photon-drag effects [60,61].

**Acknowledgment:**

R. F. is supported by the National Natural Science Foundation of China Grant No. 12204035 and the Air Force Office of Scientific Research (AFOSR) Grant No. FA9550-20-1-0255, and L.Y. is supported by the National Science Foundation (NSF) Grant No. DMR-2124934. The computational resources are provided by the Extreme Science and Engineering Discovery Environment (XSEDE), which is supported by National Science Foundation (NSF) Grant No. ACI-1548562. The authors acknowledge the Texas Advanced Computing Center (TACC) at The University of Texas at Austin for providing HPC resources.

Email: *rfei@bit.edu.cn
†lyang@physics.wustl.edu

**Figures:**

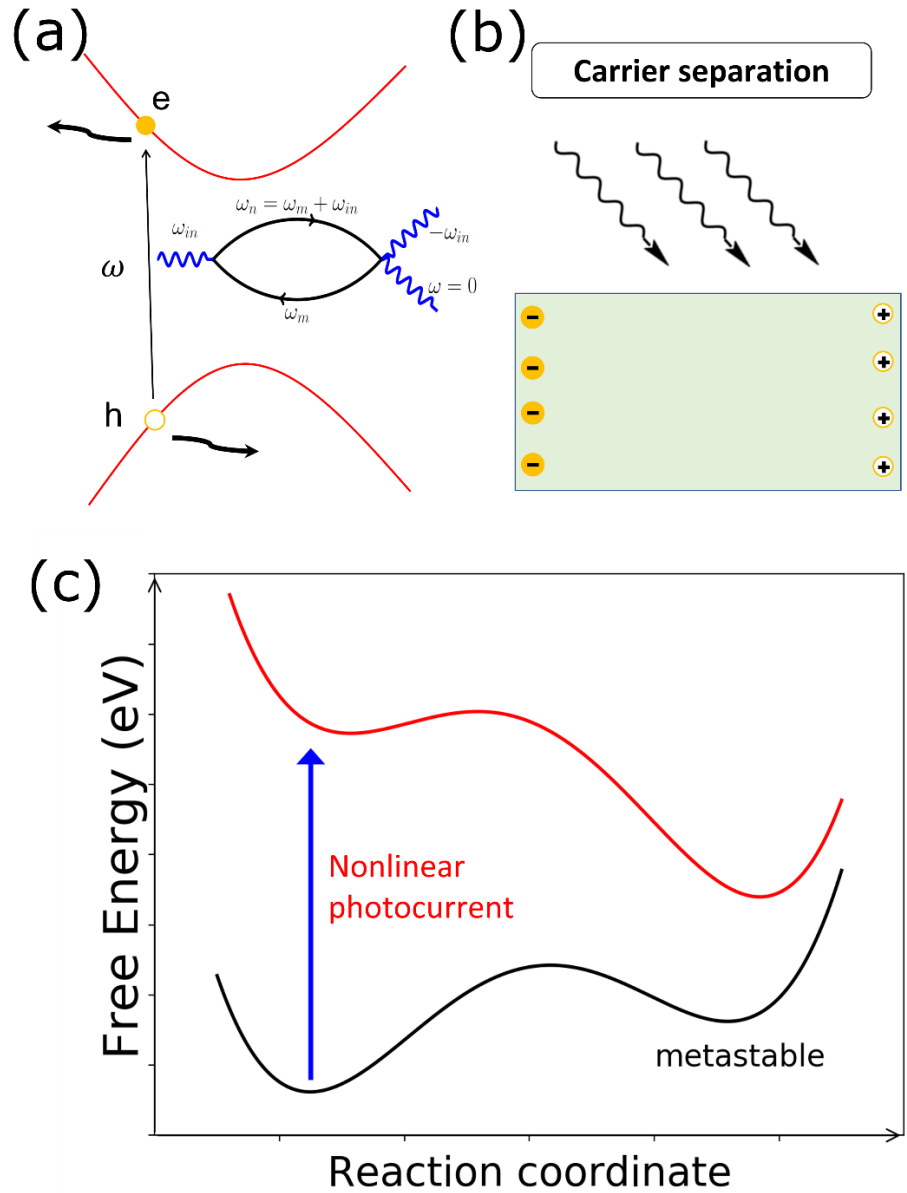

FIG. 1. (a) Nonlinear optical response induced the carrier separation in homogeneous materials, after photoexcitation. The insert Feynman diagram represents the shift current, a second-order photocurrent under linearly polarized light in nonmagnetic materials. (b) In a classical depiction, a capacitor is formed by the separation of electrons and holes on opposite sides in material under the open-circuit situation. (c) Phenomenologically, the acquisition of energy through charging the capacitor raises the energy potential of structures, resulting in the photostrictive effect or structural phase transition.

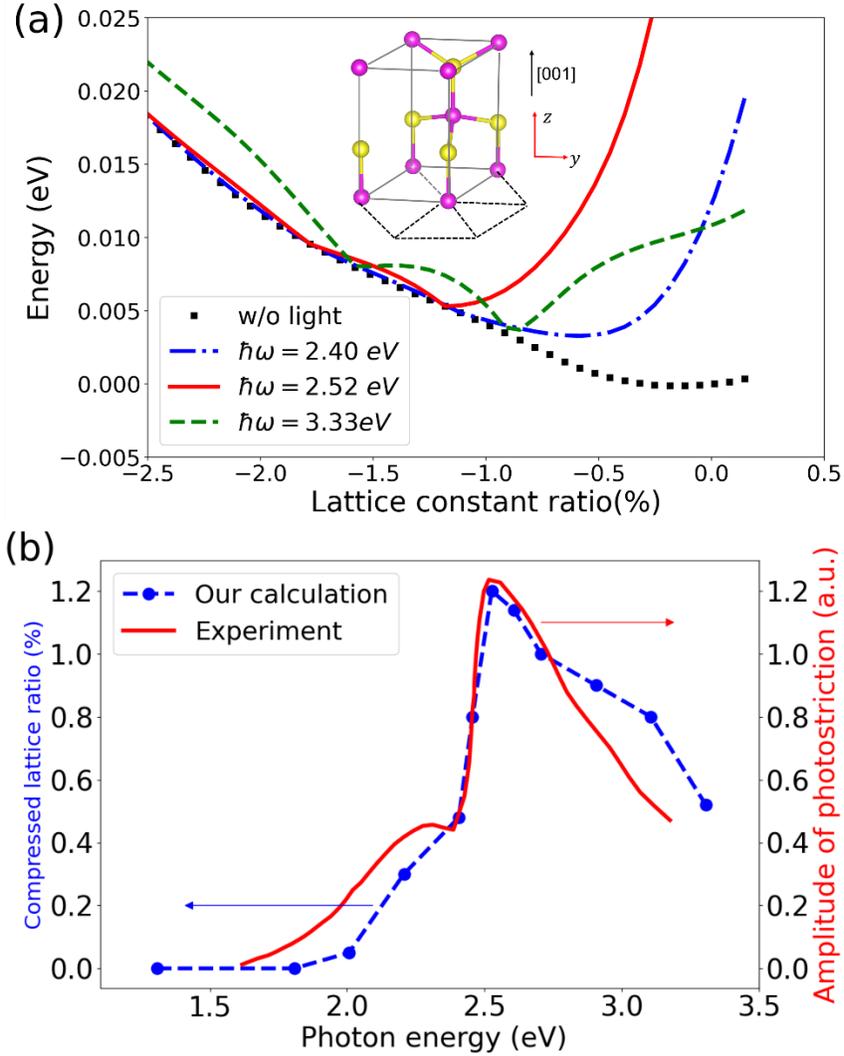

FIG. 2. (a) Total energy vs lattice constant ($z$-axis direction) of intrinsic wurtzite CdS and its structure irradiated by different photon energies. The insert figure is the atomic structure of wurtzite CdS. (b) Comparison between calculated lattice compression percentage (blue dashed line) and the experimental findings from Ref. [9] (red line) for different frequency photons. The unit of measurement for the photostriction amplitude in Ref. [9] is in arbitrary units.

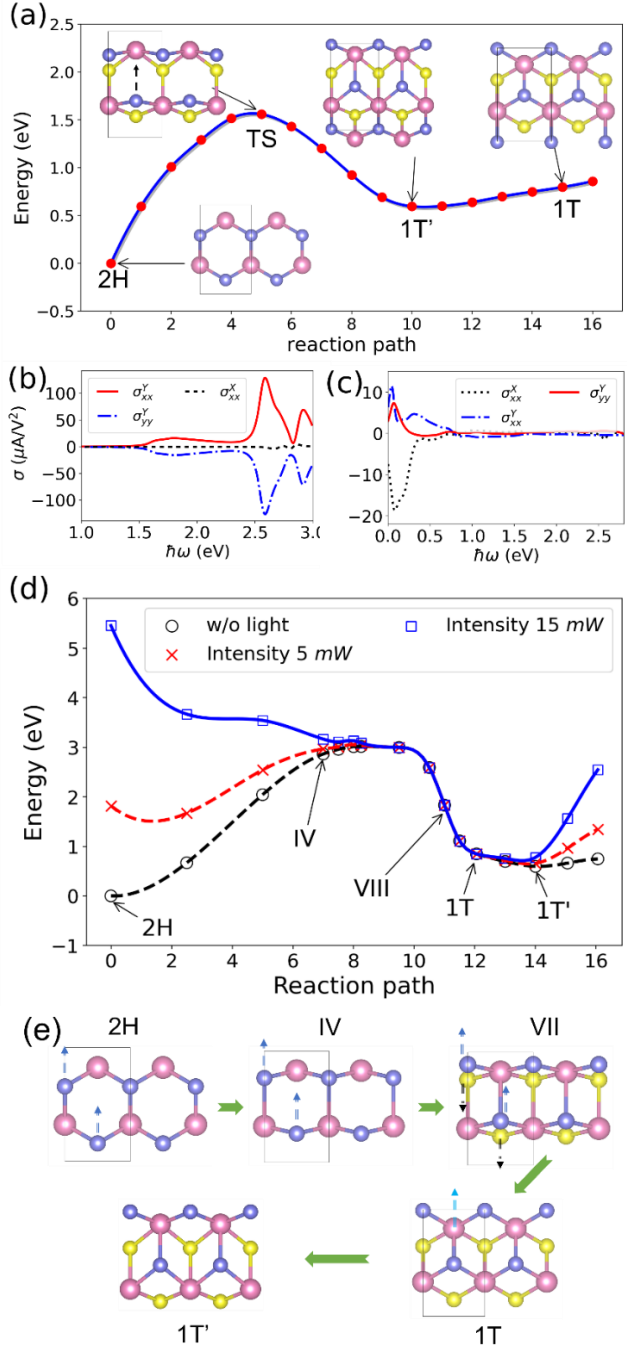

FIG. 3. (a) Relative total energy per $MoS_2$ unit cell as a function of the fractional reaction coordinate. (b) and (c) Shift current susceptibility of the 2H phase and 1T' phase, respectively. (d) The laser-induced increased potential of monolayer $MoS_2$. The circle dot represents the potential of the new reaction path indicated by (e) without laser irradiation, the cross dot and square dot represent the structures irradiated by $1 \mu m$ diameter laser beam with $5 mW$ power and $15 mW$ power, respectively. The average irradiation time on the sample is $10 s$ in this calculation.

**Reference:**
[1] A. de la Torre, D. M. Kennes, M. Claassen, S. Gerber, J. W. McIver, and M. A. Sentef, Colloquium: *Nonthermal Pathways to Ultrafast Control in Quantum Materials*, Rev Mod Phys **93**, 041002 (2021).
[2] D. Fausti, R. I. Tobey, N. Dean, S. Kaiser, A. Dienst, M. C. Hoffmann, S. Pyon, T. Takayama, H. Takagi, and A. Cavalleri, *Light-Induced Superconductivity in a Stripe-Ordered Cuprate*, Science **331**, 189 (2011).
[3] A. Cavalleri, Cs. Tóth, C. W. Siders, J. A. Squier, F. Ráksi, P. Forget, and J. C. Kieffer, *Femtosecond Structural Dynamics in VO2 during an Ultrafast Solid-Solid Phase Transition*, Phys Rev Lett **87**, 237401 (2001).
[4] E. Collet et al., *Laser-Induced Ferroelectric Structural Order in an Organic Charge-Transfer Crystal*, Science **300**, 612 (2003).
[5] X. Li, T. Qiu, J. Zhang, E. Baldini, J. Lu, A. M. Rappe, and K. A. Nelson, *Terahertz Field–Induced Ferroelectricity in Quantum Paraelectric $SrTiO_3$*, Science **364**, 1079 (2019).
[6] T. F. Nova, A. S. Disa, M. Fechner, and A. Cavalleri, *Metastable Ferroelectricity in Optically Strained $SrTiO_3$*, Science **364**, 1075 (2019).
[7] Y. H. Wang, H. Steinberg, P. Jarillo-Herrero, and N. Gedik, *Observation of Floquet-Bloch States on the Surface of a Topological Insulator*, Science **342**, 453 (2013).
[8] J. W. McIver, B. Schulte, F.-U. Stein, T. Matsuyama, G. Jotzu, G. Meier, and A. Cavalleri, *Light-Induced Anomalous Hall Effect in Graphene*, Nat Phys **16**, 38 (2020).
[9] J. Lagowski and H. C. Gatos, *Photomechanical Effect in Noncentrosymmetric Semiconductors-CdS*, Appl Phys Lett **20**, 14 (1972).
[10] B. Kundys, M. Viret, D. Colson, and D. O. Kundys, *Light-Induced Size Changes in BiFeO3 Crystals*, Nat Mater **9**, 803 (2010).
[11] B. Kundys, *Photostrictive Materials*, Appl Phys Rev **2**, 011301 (2015).
[12] Y. Zhou et al., *Giant Photostriction in Organic–Inorganic Lead Halide Perovskites*, Nat Commun **7**, 11193 (2016).
[13] S. Cho et al., *Phase Patterning for Ohmic Homojunction Contact in MoTe2*, Science **349**, 625 (2015).
[14] M. Y. Zhang et al., *Light-Induced Subpicosecond Lattice Symmetry Switch in MoTe2*, Phys Rev X **9**, 021036 (2019).
[15] H. Mine et al., *Laser-Beam-Patterned Topological Insulating States on Thin SemiconductingMoS2.*, Phys Rev Lett **123**, 146803 (2019).
[16] A. Gómez-León and G. Platero, *Floquet-Bloch Theory and Topology in Periodically Driven Lattices*, Phys Rev Lett **110**, 200403 (2013).
[17] M. Nuske, L. Broers, B. Schulte, G. Jotzu, S. A. Sato, A. Cavalleri, A. Rubio, J. W. McIver, and L. Mathey, *Floquet Dynamics in Light-Driven Solids*, Phys Rev Res **2**, 043408 (2020).
[18] T. Morimoto and N. Nagaosa, *Topological Nature of Nonlinear Optical Effects in Solids*, Sci Adv **2**, e1501524 (2016).
[19] B. Dóra, J. Cayssol, F. Simon, and R. Moessner, *Optically Engineering the Topological Properties of a Spin Hall Insulator*, Phys Rev Lett **108**, 056602 (2012).
[20] S. Meng and E. Kaxiras, *Real-Time, Local Basis-Set Implementation of Time-Dependent Density Functional Theory for Excited State Dynamics Simulations*, J Chem Phys **129**, 054110 (2008).

[21] H. O. Wijewardane and C. A. Ullrich, *Real-Time Electron Dynamics with Exact-Exchange Time-Dependent Density-Functional Theory*, Phys Rev Lett **100**, 056404 (2008).
[22] Z. Wang, S.-S. Li, and L.-W. Wang, *Efficient Real-Time Time-Dependent Density Functional Theory Method and Its Application to a Collision of an Ion with a 2D Material*, Phys Rev Lett **114**, 063004 (2015).
[23] K. Yabana and G. F. Bertsch, *Time-Dependent Local-Density Approximation in Real Time*, Phys Rev B **54**, 4484 (1996).
[24] P. Tangney and S. Fahy, *Calculations of the A1 Phonon Frequency in Photoexcited Tellurium*, Phys Rev Lett **82**, 4340 (1999).
[25] C. Paillard, E. Torun, L. Wirtz, J. Íñiguez, and L. Bellaiche, *Photoinduced Phase Transitions in Ferroelectrics*, Phys Rev Lett **123**, 087601 (2019).
[26] L. Gao, C. Paillard, and L. Bellaiche, *Photoinduced Control of Ferroelectricity in Hybrid-Improper Ferroelectric Superlattices*, Phys Rev B **107**, 104109 (2023).
[27] B. Kundys, M. Viret, C. Meny, V. Da Costa, D. Colson, and B. Doudin, *Wavelength Dependence of Photoinduced Deformation in BiFeO3*, Phys Rev B **85**, 092301 (2012).
[28] W. T. H. Koch, R. Munser, W. Ruppel, and P. Würfel, *Anomalous Photovoltage in BaTiO 3*, Ferroelectrics **13**, 305 (1976).
[29] T. Choi, S. Lee, Y. J. Choi, V. Kiryukhin, and S.-W. Cheong, *Switchable Ferroelectric Diode and Photovoltaic Effect in BiFeO3*, Science **324**, 63 (2009).
[30] S. Y. Yang et al., *Above-Bandgap Voltages from Ferroelectric Photovoltaic Devices*, Nat Nanotechnol **5**, 143 (2010).
[31] H. Wang and X. Qian, *Ferroicity-Driven Nonlinear Photocurrent Switching in Time-Reversal Invariant Ferroic Materials*, Sci Adv **5**, eaav9743 (2019).
[32] M. Sotome, M. Nakamura, T. Morimoto, Y. Zhang, G.-Y. Guo, M. Kawasaki, N. Nagaosa, Y. Tokura, and N. Ogawa, *Terahertz Emission Spectroscopy of Ultrafast Exciton Shift Current in the Noncentrosymmetric Semiconductor CdS*, Phys Rev B **103**, L241111 (2021).
[33] H. Yuan et al., *Generation and Electric Control of Spin–Valley-Coupled Circular Photogalvanic Current in WSe2*, Nat Nanotechnol **9**, 851 (2014).
[34] J. Łagowski and H. C. Gatos, *Photomechanical Vibration of Thin Crystals of Polar Semiconductors*, Surf Sci **45**, 353 (1974).
[35] T. Figielski, *Photostriction Effect in Germanium*, Physica Status Solidi (b) **1**, 306 (1961).
[36] R. von Baltz and W. Kraut, *Theory of the Bulk Photovoltaic Effect in Pure Crystals*, Phys Rev B **23**, 5590 (1981).
[37] J. E. Sipe and A. I. Shkrebtii, *Second-Order Optical Response in Semiconductors*, Phys Rev B **61**, 5337 (2000).
[38] S. D. Ganichev, H. Ketterl, W. Prettl, E. L. Ivchenko, and L. E. Vorobjev, *Circular Photogalvanic Effect Induced by Monopolar Spin Orientation in P-GaAs/AlGaAs Multiple-Quantum Wells*, Appl Phys Lett **77**, 3146 (2000).
[39] Y. Zhang, T. Holder, H. Ishizuka, F. de Juan, N. Nagaosa, C. Felser, and B. Yan, *Switchable Magnetic Bulk Photovoltaic Effect in the Two-Dimensional Magnet CrI3*, Nat Commun **10**, 3783 (2019).
[40] R. Fei, W. Song, and L. Yang, *Giant Photogalvanic Effect and Second-Harmonic Generation in Magnetic Axion Insulators*, Phys Rev B **102**, 035440 (2020).
[41] Z. Dai, A. M. Schankler, L. Gao, L. Z. Tan, and A. M. Rappe, *Phonon-Assisted Ballistic Current from First-Principles Calculations*, Phys Rev Lett **126**, 177403 (2021).

[42] H. Watanabe and Y. Yanase, *Chiral Photocurrent in Parity-Violating Magnet and Enhanced Response in Topological Antiferromagnet*, Phys Rev X **11**, 011001 (2021).
[43] L. E. Golub and E. L. Ivchenko, *Circular and Magnetoinduced Photocurrents in Weyl Semimetals*, Phys Rev B **98**, 075305 (2018).
[44] S. M. Young and A. M. Rappe, *First Principles Calculation of the Shift Current Photovoltaic Effect in Ferroelectrics*, Phys Rev Lett **109**, 116601 (2012).
[45] D. E. Parker, T. Morimoto, J. Orenstein, and J. E. Moore, *Diagrammatic Approach to Nonlinear Optical Response with Application to Weyl Semimetals*, Phys Rev B **99**, 045121 (2019).
[46] See Supplemental Material at XX, which includes Refs. 47-50. We show the first-principles computational details, derivation of the light-injected energy in CdS and MoS2, and calculated results using different methods and packages.
[47] G. Kresse and J. Furthmüller, *Efficient Iterative Schemes for Ab Initio Total-Energy Calculations Using a Plane-Wave Basis Set*, Phys Rev B **54**, 11169 (1996).
[48] G. Kresse and D. Joubert, *From Ultrasoft Pseudopotentials to the Projector Augmented-Wave Method*, Phys Rev B **59**, 1758 (1999).
[49] J. P. Perdew, K. Burke, and M. Ernzerhof, *Generalized Gradient Approximation Made Simple*, Phys Rev Lett **77**, 3865 (1996).
[50] P. Giannozzi et al., *QUANTUM ESPRESSO: A Modular and Open-Source Software Project for Quantum Simulations of Materials*, Journal of Physics: Condensed Matter **21**, 395502 (2009).
[51] Otfried Madelung, *Semiconductors: Data Handbook* (Springer Berlin, Heidelberg, n.d.).
[52] J. Shi et al., *Terahertz-Field-Driven Metastable Topological Phase in an Atomically Thin Crystal*, ArXiv arXiv:1910.13609 (2019).
[53] X. Qian, J. Liu, L. Fu, and J. Li, *Quantum Spin Hall Effect in Two-Dimensional Transition Metal Dichalcogenides*, Science **346**, 1344 (2014).
[54] J. Jiang, Z. Chen, Y. Hu, Y. Xiang, L. Zhang, Y. Wang, G.-C. Wang, and J. Shi, *Flexo-Photovoltaic Effect in MoS2*, Nat Nanotechnol **16**, 894 (2021).
[55] S. R. Panday, S. Barraza-Lopez, T. Rangel, and B. M. Fregoso, *Injection Current in Ferroelectric Group-IV Monochalcogenide Monolayers*, Phys Rev B **100**, 195305 (2019).
[56] S. O. Mariager et al., *Structural and Magnetic Dynamics of a Laser Induced Phase Transition in FeRh*, Phys Rev Lett **108**, 087201 (2012).
[57] C. Zhang, P. Guo, and J. Zhou, *Tailoring Bulk Photovoltaic Effects in Magnetic Sliding Ferroelectric Materials*, Nano Lett **22**, 9297 (2022).
[58] A. Kogar et al., *Light-Induced Charge Density Wave in LaTe3*, Nat Phys **16**, 159 (2020).
[59] L. Luo et al., *A Light-Induced Phononic Symmetry Switch and Giant Dissipationless Topological Photocurrent in ZrTe5*, Nat Mater **20**, 329 (2021).
[60] D. R. Ward, F. Hüser, F. Pauly, J. C. Cuevas, and D. Natelson, *Optical Rectification and Field Enhancement in a Plasmonic Nanogap*, Nat Nanotechnol **5**, 732 (2010).
[61] J. H. Strait, G. Holland, W. Zhu, C. Zhang, B. R. Ilic, A. Agrawal, D. Pacifici, and H. J. Lezec, *Revisiting the Photon-Drag Effect in Metal Films*, Phys Rev Lett **123**, 053903 (2019).